\documentclass[twocolumn,english,aps,prl,superscriptaddress]{revtex4-1}
\usepackage{amsmath}
\usepackage{newtxmath}
\usepackage[latin9]{inputenc}
\usepackage{mathrsfs}
\usepackage{graphicx}

\makeatletter
\usepackage[colorlinks,citecolor=blue,linkcolor=blue]{hyperref}

\makeatother

\usepackage{babel}
\begin{document}

\title{Directly profiling the dark-state transition density via scanning
tunneling microscope}

\author{Guohui Dong}

\affiliation{Graduate School of Chinese Academy of Engineering Physics, Beijing
100084, China}

\author{Zhubin Hu}

\affiliation{New York University Shanghai and NYU-ECNU Center for Computational
Chemistry, 1555 Century Avenue, Shanghai 200122, China}

\author{Xiang Sun}
\email{xiang.sun@nyu.edu}

\affiliation{New York University Shanghai and NYU-ECNU Center for Computational
Chemistry, 1555 Century Avenue, Shanghai 200122, China}

\affiliation{Department of Chemistry, New York University, New York, New York
10003, United States}

\author{Hui Dong}
\email{hdong@gscaep.ac.cn}

\affiliation{Graduate School of Chinese Academy of Engineering Physics, Beijing
100084, China}
\begin{abstract}
The molecular dark state participates in many important photon-induced
processes, yet is typically beyond the optical-spectroscopic measurement
due to the forbidden transition dictated by the selection rule. In this
work, we propose to use the scanning tunneling microscope (STM) as
an incisive tool to directly profile the dark-state transition density
of a single molecule, taking advantage of the localized static electronic
field near the metal tip. The detection of dark state is achieved by
measuring the fluorescence from a higher bright state to the ground state with assistant optical
pumping. The current proposal shall bring new methodology to study
the single-molecule properties in the electro-optical devices and
the light-assisted biological processes.
\end{abstract}
\maketitle

\textit{Introduction} -- Controllable light-matter interaction in the
nanometer scale is one of the most fundamental and attractive topics in
areas such as laser techniques \citep{Haken_laser_theory,Scully_quantum_optics},
atom manipulation \citep{1987_Steven_Chu_MOT,1988_Phillips_MOT,MOT_1990_Cohen_Tannoudji,atom_manipulation_1999},
and cavity quantum electrodynamics \citep{cavity_QED_Dutra_2005,cavity_QED_2006,Agarwal_quantum_optics_2013}.
Typically, the wavelength of the optical field is several orders of magnitude
larger than the size of the matter of interest. In this region, the
systems are essentially manipulated under the dipole interaction $\vec{\mathbf{\mu}}\cdot\vec{E}$,
where $\vec{\mathbf{\mu}}$ is the electric dipole of the matter and $\vec{E}$ is the external electric field.
Consequently, only the transitions between the atomic or molecular states with nonzero
transition dipoles can be probed with electromagnetic field, leaving
the dark state with zero transition dipole beyond the optical detection.
For this reason, traditional optical spectroscopic approaches such as infrared \citep{IR_2004}, Raman \citep{IR&Raman_2011},
and fluorescence \citep{fluorescence_2000,fluorescence_2010} spectroscopies
are not applicable for the detection of dark states.
However, the molecular dark state
plays a significant role in many biochemical processes, such as it helps to resist the photochemical damage to the deoxyribonucleic acid (DNA) induced by ultraviolet light
\citep{2009_DNA_damage} and assist in the energy transfer process
in the photosynthetic systems \citep{2018_photosynthesis,2017_photosynthesis}.

The straightforward method is to break the dipole approximation with the spatial modulated field on the scale comparable to a single molecule.
Such modulated field can be found near the tip of the scanning tunneling microscope (STM), which is as small as several atoms and able to induce electronic excitation which would have been forbidden under the dipole approximation.
In contrast to optical excitation, the electronic excitation induced in STM provides
detailed information on molecular states \citep{Repp_2005,Ho_2010}.
By counting the luminescence photon, the scanning-tunneling-microscope-induced
luminescence (STML) has emerged as a crucial tool for studying photoelectronic properties of single molecules \citep{Ros_awska_2018,Nian_simulation_2018,HiroshiImada_fano_PRL_2017,Doppagne_science_2018,GongChen_PRL_2019_triplet_up_conversion,Kroger_nanolett_fano_2018,simulation_xyWu_2019}.

In this Letter, we show the scheme of directly profiling the dark-state transition
details from STM \citep{2020_GHDong_STML}
taking advantage of the localized electric field. We quantitatively demonstrate the
resemblance between the relative inelastic current (luminescence photon
counting) and the transition density profile of the dark-state transition.

\begin{figure}[t]
\begin{centering}
\includegraphics[scale=0.6]{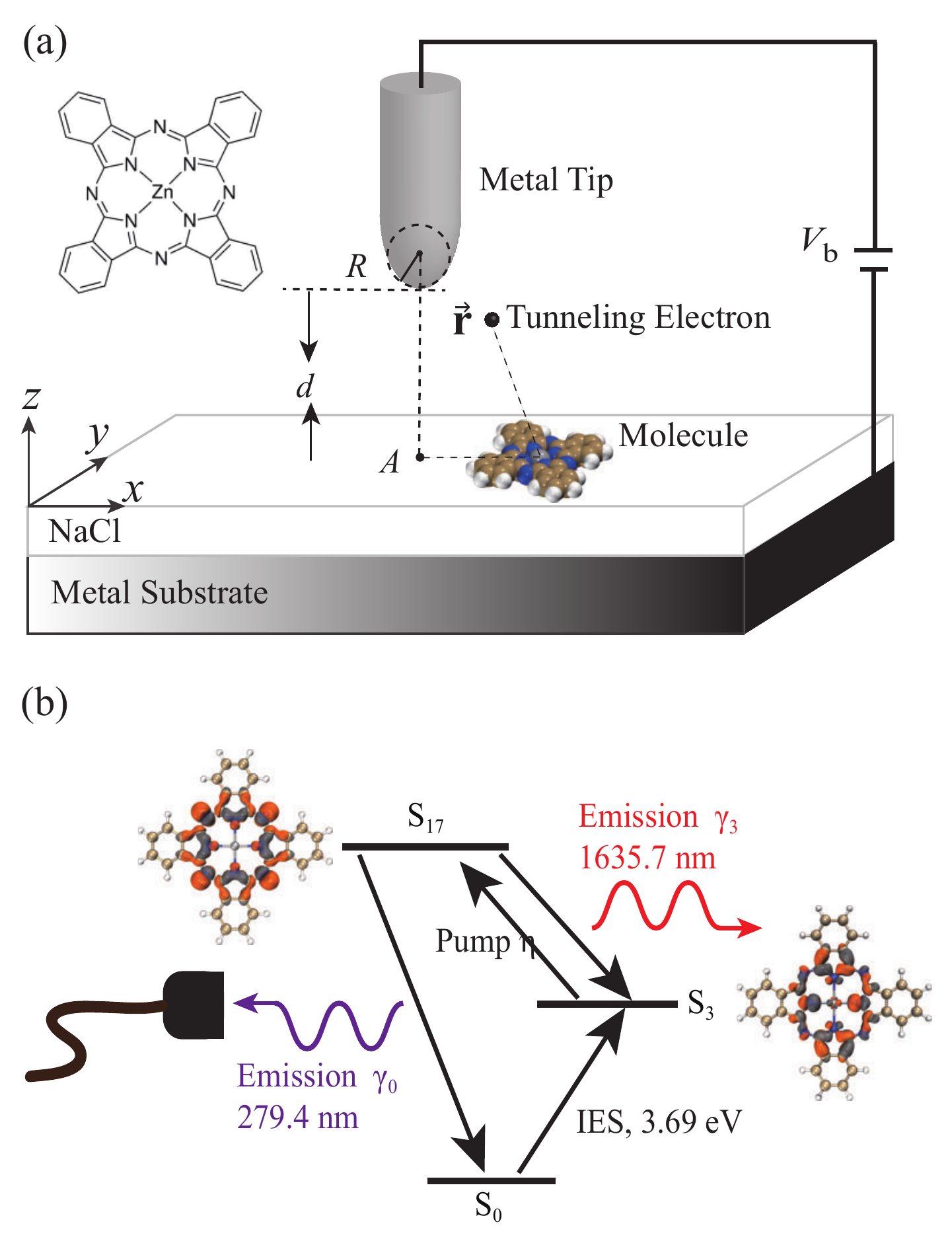}
\par\end{centering}
\caption{\label{fig:schematic_diagrame}(Color online) (a) Schematic diagram
of scanning-tunneling-microscope-induced
luminescence (STML). A single molecule is placed on a NaCl-covered metal plane.
The STM tip apex is treated as a sphere with radius $R$. Point $A$
represents the projection position of tip's center on the plane, and
$d$ stands for the distance between the tip and plane. The tunneling
electron is shown as a black sphere with vector $\vec{\mathbf{r}}$,
and the center position of the molecule is set as the origin of the
reference frame. Inset: ZnPc molecule. (b) Schematic diagram of the
detection of dark state $\mathrm{S}_{3}$. Both the transition dipole
moments between $\mathrm{S}_{0}$-$\mathrm{S}_{17}$ and $\mathrm{S}_{3}$-$\mathrm{S}_{17}$
are nonzero. Firstly the molecule is excited to the dark state through
the IES process and a laser with the wavelength $1635.7\mathrm{nm}$
couples the dark state $\mathrm{S}_{3}$ and the bright state $\mathrm{S}_{17}$.
The molecule in $\mathrm{S}_{17}$ emits photon through decaying to
$\mathrm{S}_{0}$ and $\mathrm{S}_{3}$. Photons at the wavelength
$279.4\mathrm{nm}$ are collected to profile the dark state. Insets:
the transition density of transitions $\mathrm{S}_{0}$-$\mathrm{S}_{17}$
and $\mathrm{S}_{0}$-$\mathrm{S}_{3}$. The orange and gray parts
represent the positive and negative transition density, respectively.}
\end{figure}

\textit{Model} --We demonstrate the basic setup of the model in Fig.
\ref{fig:schematic_diagrame}(a), where a single molecule is placed
on a NaCl-covered metal substrate. A metal tip scans over the molecule
to allow the profiling. Driven at a nonzero bias voltage, an electron
tunnels from one electrode to the other, while interacting with the
molecule through the Coulomb interaction.

The total Hamiltonian consists of the tunneling electron Hamiltonian
$\hat{H}_{\mathrm{el}}$, the molecular Hamiltonian $\hat{H}_{\mathrm{m}}$,
and the electron-molecule interaction Hamiltonian $\hat{H}_{\mathrm{el-m}}$.
The tunneling electron Hamiltonian is $\hat{H}_{\mathrm{el}}=-\hat{\nabla}^{2}/(2m_{e})+V(\hat{\vec{\mathbf{r}}})$, where
$V(\hat{\vec{\mathbf{r}}})$ stands for the tunneling-electron potential
at $\hat{\vec{\mathbf{r}}}=(x,y,z)$ and $m_{e}$ is the electron
mass. The free tip and substrate  wavefunctions are written as \citep{bardeen_1961,TUTORIAL_2006,Tersoff_Hamann_1983,Tersoff_Hamann_1985,2020_GHDong_STML}

\begin{subequations}
\begin{align}
\hat{H}_{\mathrm{el,t}}\left|\phi_{k}\right\rangle  & \simeq\widetilde{\xi}_{k}\left|\phi_{k}\right\rangle ,\label{eq:solution_tip=000026base_nonzero_bias}\\
\hat{H}_{\mathrm{el,s}}\left|\varphi_{n}\right\rangle  & \simeq\widetilde{E}_{n}\left|\varphi_{n}\right\rangle,
\end{align}
\end{subequations}
respectively. Here, $\hat{H}_{\mathrm{el,t}}$ ($\hat{H}_{\mathrm{el,s}}$)
is the free tip (substrate) Hamiltonian without the corresponding potential in the
substrate (tip) region. $\left|\phi_{k}\right\rangle \left(\left|\varphi_{n}\right\rangle \right)$
is the eigenfunction with eigenenergy $\widetilde{\xi}_{k}\equiv\xi_{k}+eV_{\mathrm{b}}\left(\widetilde{E}_{n}\equiv E_{n}\right)$,
and $\xi_{k}\left(E_{n}\right)$ is its corresponding eigenenergy
at zero bias voltage \citep{2020_GHDong_STML}. The molecular Hamiltonian
is simplified as a multi-level system $\hat{H}_{\mathrm{m}}=E_{g}\left|\chi_{g}\right\rangle \left\langle \chi_{g}\right|+\sum_{i=1}^{l}E_{e,i}\left|\chi_{e,i}\right\rangle \left\langle \chi_{e,i}\right|$,
where $|\chi_{g}\rangle\:(|\chi_{e,i}\rangle)$ is its ground ($i$-th
excited) state with energy $E_{g}\left(E_{e,i}\right)$ and $l$ is
the total number of excited states.

The excitation of the molecules are performed through the electron-molecule
interaction $\hat{H}_{\mathrm{el-m}}$ \citep{2020_GHDong_STML} as

\begin{align}
\hat{H}_{\mathrm{el-m}} & \simeq-\sum_{i=1}^{l}\sum_{n,k}\mathscr{N}_{\mathrm{s,t};i}|_{V_{\mathrm{b}},E_{n}\rightarrow\xi_{k}}\left|\phi_{k}\right\rangle \left\langle \varphi_{n}\right|\otimes\hat{\sigma}_{x,i}\nonumber \\
 & \simeq-\sum_{i=1}^{l}\sum_{n,k}\mathscr{N}_{\mathrm{t,s};i}|_{V_{\mathrm{b}},\xi_{k}\rightarrow E_{n}}\left|\varphi_{n}\right\rangle \left\langle \phi_{k}\right|\otimes\hat{\sigma}_{x,i},\label{eq:interaction_H}
\end{align}
where the transition matrix element

\begin{align}
\mathscr{N}_{\mathrm{s,t};i}|_{V_{\mathrm{b}},E_{n}\rightarrow\xi_{k}} & =\mathscr{N}_{\mathrm{t,s};i}|_{V_{\mathrm{b}},\xi_{k}\rightarrow E_{n}}\nonumber \\
 & =e^{2}\int d^{3}\vec{q}\rho_{T,i}\left(\vec{q}\right)\rho_{n,k}\left(-\vec{q}\right)\frac{4\pi}{q^{2}}\label{eq:N}
\end{align}
describes the transition matrix element from $\left|\phi_{k}\right\rangle \left(\left|\varphi_{n}\right\rangle \right)$
to $\left|\varphi_{n}\right\rangle \left(\left|\phi_{k}\right\rangle \right)$
in the subspace $\left|\chi_{g}\right\rangle ,\left|\chi_{e,i}\right\rangle $,
and $\hat{\sigma}_{x,i}\equiv\left|\chi_{e,i}\right\rangle \left\langle \chi_{g}\right|+\left|\chi_{g}\right\rangle \left\langle \chi_{e,i}\right|$
stands for the molecular transition. Here $\rho_{T,i}\left(\vec{q}\right)$
and $\rho_{n,k}\left(\vec{q}\right)$ are the Fourier transforms of
the transition density $\rho_{T,i}\left(\vec{r}\right)\equiv\left\langle \vec{r}|\chi_{e,i}\right\rangle \left\langle \chi_{g}|\vec{r}\right\rangle $
and the product of the wavefunctions of the tip and the substrate
$\rho_{n,k}\left(\vec{r}\right)\equiv\left\langle \vec{r}|\phi_{k}\right\rangle \left\langle \varphi_{n}|\vec{r}\right\rangle $,
respectively. The transition dipole moment $\boldsymbol{\vec{\mu}}_{i}$
is obtained as the integral of the transition density $\rho_{T,i}\left(\vec{r}\right)$
and vector $\vec{r}$, i.e., $\boldsymbol{\vec{\mu}}_{i}=\int d^{3}\vec{r}\rho_{T,i}\left(\vec{r}\right)\vec{r}$.
Detailed derivation is provided in  Supplementary Material. Beyond
the dipole approximation, the transition
matrix element here  is expressed as a convolution of the Fourier
transform of the transition density and the electrode wavefunctions.

The properties of the molecules can be  described by the tunneling
current and the photon counting. The tunneling current is calculated
to the first order ($\hat{H}_{\mathrm{el}}-\hat{H}_{\mathrm{el,s}}$ and
$\hat{H}_{\mathrm{el-m}}$ as the perturbation). At the negative bias
$V_{\mathrm{b}}<0$, the molecule is initially in its ground state
and the electron is in the substrate eigenstate, i.e., $\left|\Psi\left(t=0\right)\right\rangle =\left|\chi_{g}\right\rangle \left|\varphi_{n}\right\rangle $.
The wavefunction at time $t$ evolves as
\begin{align}
\left|\Psi\left(t\right)\right\rangle  & =e^{-i\left(\widetilde{E}_{n}+E_{g}\right)t}\left|\chi_{g}\right\rangle \left|\varphi_{n}\right\rangle \nonumber \\
 & +\sum_{k}\left[c_{g;k}\left(t\right)\left|\chi_{g}\right\rangle +\sum_{i=1}^{l}c_{e,i;k}\left(t\right)\left|\chi_{e,i}\right\rangle \right]\left|\phi_{k}\right\rangle .
\end{align}
Here $c_{g;k}\left(t\right)$ shows the probability amplitude of the
elastic tunneling process and $c_{e,i;k}\left(t\right)$ the probability
amplitude of the inelastic tunneling process. In the rotating-wave
approximation, the inelastic tunneling amplitude becomes
\[
c_{e,i;k}\left(t\right)=\frac{e^{-i\left(\widetilde{E}_{n}+E_{g}\right)t}-e^{-i\left(\widetilde{\xi}_{k}+E_{e,i}\right)t}}{\widetilde{E}_{n}-\widetilde{\xi}_{k}-E_{eg,i}}\mathscr{N}_{\mathrm{s,t};i}|_{V_{\mathrm{b}},E_{n}\rightarrow\xi_{k}},
\]
where $E_{eg,i}\equiv E_{e,i}-E_{g}$ is the energy gap between the
molecular states $\left|\chi_{g}\right\rangle $ and $\left|\chi_{e,i}\right\rangle $.
By tracing out the degrees of freedom of the molecule, we obtain the
inelastic current from $\left|\varphi_{n}\right\rangle $ to $\left|\phi_{k}\right\rangle $
as $\mathscr{J}_{n\rightarrow k}=\sum_{i=1}^{l}d\left|c_{e,i;k}\left(t\right)\right|^{2}/dt$
and the total inelastic current as
\begin{align}
I_{-\mathrm{,inela}} & \simeq\sum_{i=1}^{l}I_{-\mathrm{,inela},i},\label{eq:inela_current}
\end{align}
where
\begin{align}
I_{-\mathrm{,inela},i} & =2\pi e\int_{\mu_{0}+eV_{\mathrm{b}}+E_{eg,i}}^{\mu_{0}}dE_{n}\rho_{\mathrm{s}}\left(E_{n}\right)\rho_{\mathrm{t}}\left(\xi_{k}\right)\nonumber \\
 & \times\left|\mathscr{N}_{\mathrm{s,t};i}|_{V_{\mathrm{b}},E_{n}\rightarrow\xi_{k}}\right|^{2}|_{\xi_{k}=E_{n}-eV_{\mathrm{b}}-E_{eg,i}}.\label{eq:inela_current_i}
\end{align}
Here $\mu_{0}$ is the Fermi energy of the electrode and $\rho_{\mathrm{t}}\left(E\right)$
($\rho_{\mathrm{s}}\left(E\right)$) is the density of state of the
tip (substrate) at  energy $E$. With the inelastic current at both
negative and positive bias, we obtain the inelastic current as below
(for the inelastic current at positive bias, see  Supplementary
Material)
\begin{equation}
I_{\mathrm{inela}}=\begin{cases}
I_{\mathrm{-,inela}}, & V_{\mathrm{b}}<-\mathrm{min}\left[\frac{E_{eg,i}}{e}\right],\\
0, & -\mathrm{min}\left[\frac{E_{eg,i}}{e}\right]\leq V_{\mathrm{b}}\leq\mathrm{min}\left[\frac{E_{eg,i}}{e}\right]\\
I_{\mathrm{+,inela}}, & V_{\mathrm{b}}>\mathrm{min}\left[\frac{E_{eg,i}}{e}\right].
\end{cases},\label{eq:I_inela}
\end{equation}
Here $\mathrm{min}\left[E_{eg,i}/e\right]$ means the minimal $E_{eg,i}/e$
for all $i$. The condition for a nonzero inelastic current is $|eV_{\mathrm{b}}|>\mathrm{min}\left[E_{eg,i}\right]$
which is a generalized result of the $l=1$ model \citep{2020_GHDong_STML}.

As shown in Eqs. (\ref{eq:N}) and (\ref{eq:inela_current_i}), the
inelastic tunneling current $I_{-\mathrm{,inela}}$ is a convolution
of the square of the molecular transition density and the wavefunctions
of the electrodes. It inherits the profile of the molecular transition
density in the $x$-$y$ plane. Thus the molecule is more likely to get excited
when the tip is above a larger transition density (absolute value).
This submolecular-resolution feature has also been captured in the
other two STML excitation mechanisms \citep{YangZhang_Nature_2016_dipole_dipole,HiroshiImada_fano_PRL_2017,simulation_xyWu_2019,2021_vibronic_coupling_Kong},
where only the bright-state excitation under the dipole approximation
is studied. Beyond the dipole approximation, our theory predicts the
dark-state excitation in the inelastic electron scattering (IES) mechanism.
This discovery will provide a new platform for the study of molecular
dark states.

\begin{figure}[t]
\begin{centering}
\includegraphics[scale=0.2]{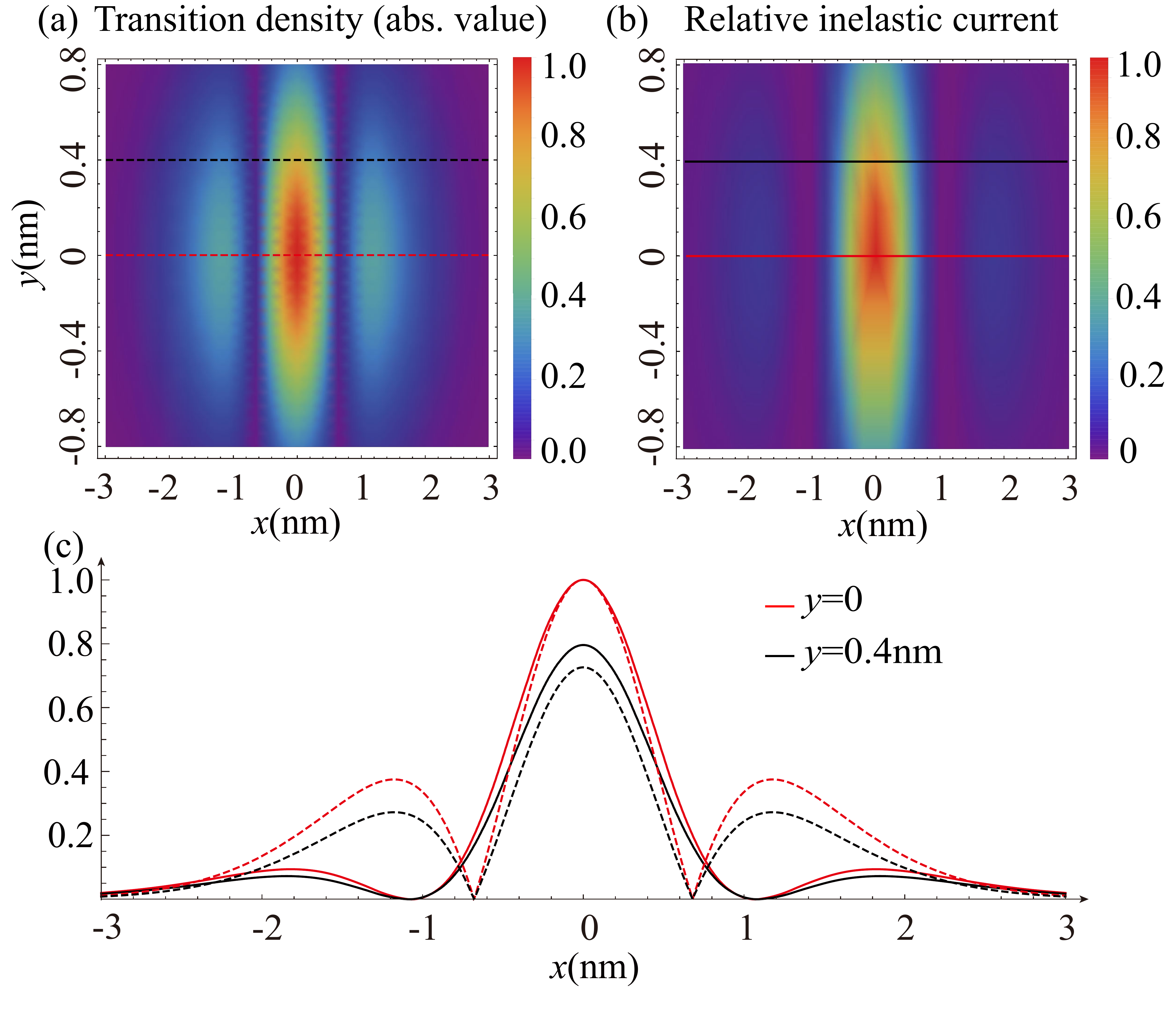}
\par\end{centering}
\caption{\label{fig:Gaussian_result-1}(Color online) (a,b) The 2D plot of
relative transition density $\left(z=0\right)$ and inelastic current
($V_{\mathrm{b}}=-2.5$V). (c) The relative transition density
(dashed lines) and the relative inelastic current (solid lines) for
$y=0,0.4\mathrm{nm}$. }
\end{figure}

\textit{General result of dark-state excitation} -- As a proof-of-principle
example, we use a simplified molecule with only two levels, namely
$l=1$. The transition density is assumed to be in a Gaussian form,

\begin{equation}
\rho_{T}\left(\vec{r}\equiv\left(x,y,z\right)\right)=\frac{1}{2\pi\sigma}\left(\frac{1}{\sigma_{1}}e^{-\frac{x^{2}}{2\sigma_{1}^{2}}}-\frac{1}{\sigma_{2}}e^{-\frac{x^{2}}{2\sigma_{2}^{2}}}\right)e^{-\frac{y^{2}}{2\sigma^{2}}}\delta\left(z\right),\label{eq:Gaussian_tran_den}
\end{equation}
where $\sigma,\sigma_{1},\sigma_{2}$ are the width of the wave packets
and $\delta\left(z\right)$ is the Dirac delta function. The transition
density is assumed in the $x$-$y$ plane, shown in Fig. \ref{fig:Gaussian_result-1}
(a)). The transition dipole is zero for the current dark state, i.e.,
$\int\vec{r}\rho_{T}(x,y,z)dxdydz=0$.

\begin{table}[t]
\begin{centering}
\includegraphics{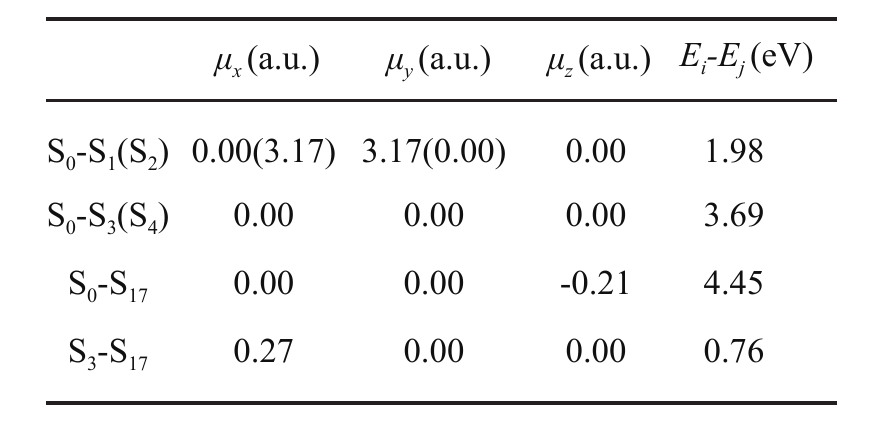}
\par\end{centering}
\caption{\label{tab:Table}The energy gap and transition dipole moments of
some electronic tansitions in the ZnPc molecule.}
\end{table}

In the calculation, we assume the silver tip and substrate with the
Fermi energy $\mu_{0}=-4.64$eV. The radius of the tip is $R=0.5$nm,
and the distance between the molecular plane and the tip is $d=1$nm.
The molecular energy gap between the ground and dark-excited state
is $E_{eg}=2$eV. Here we choose $2\sigma_{1}=\sigma_{2}=\sigma=1$nm.

Fig. \ref{fig:Gaussian_result-1} shows the transition density (subfigure
(a)) and the calculated tunneling current (subfigure (b)) from Eq.
(\ref{eq:I_inela}) with tip scanned in the $x$-$y$ plane. The bias
voltage is set as $V_{\mathrm{b}}=-2.5$V to allow a non-zero current.
The tunneling current profile resembles that of the transition density
as shown in subfigure (a) and (b). We compare the tunneling current
with the transition density at the cross section along $y=0$ and
$0.4\mathrm{nm}$ in subfigure (c). The curves show the same trend
with small deviation of the postions of the  minima.

The main features of the inelastic current in the IES mechanism in
Ref. \citep{2020_GHDong_STML} also appears in this work that goes beyond the dipole approximation. The minimal bias for nonzero inelastic current
equals the molecule  energy gap divided by the electron charge. The inelastic current at negative bias is larger than that at the
positive bias (see Supplementary Material). The two features are typical
in the IES mechanism in the cases with
or without the dipole approximation.

\begin{figure*}[t]
\begin{centering}
\includegraphics[scale=0.25]{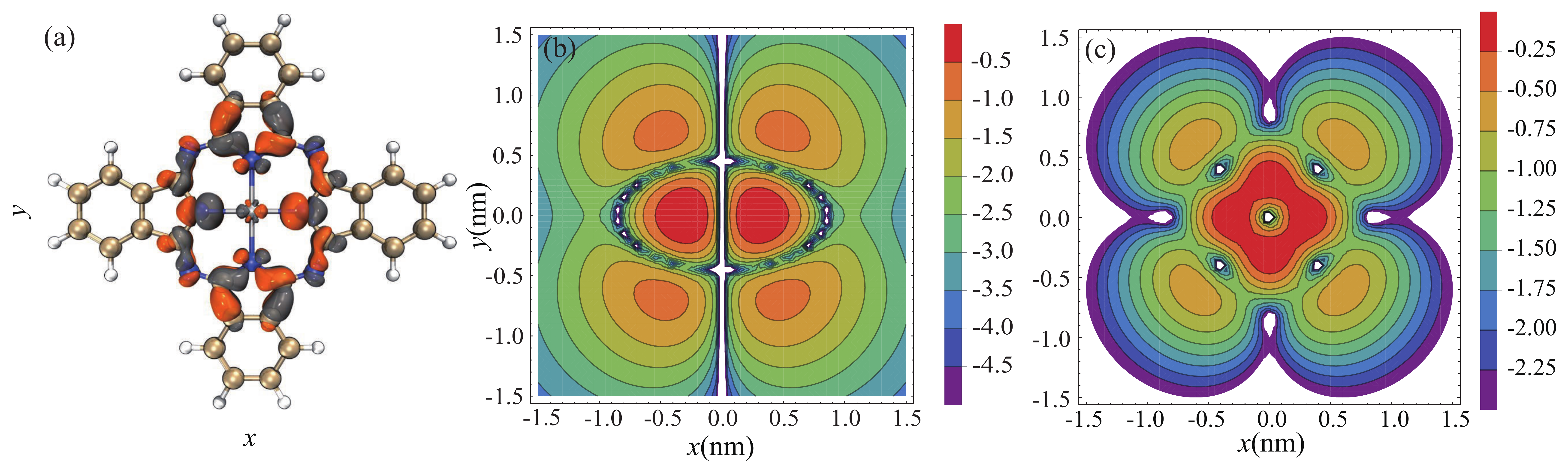}
\par\end{centering}
\centering{}\caption{\label{fig:ZnPc_tran=000026current} (a) The calculated transition
density of transition between the ground state $\mathrm{S}_{0}$ and
dark state $\mathrm{S}_{3}$. The orange and gray parts represent
the positive and negative transition density, respectively. (b) The
2D logarithmic plot of the relative inelastic current of transition
between the ground state $\mathrm{S}_{0}$ and dark state $\mathrm{S}_{3}$
at $V_{\mathrm{b}}=-4$V. (c) The 2D logarithmic plot of the sum of
the relative inelastic current of transitions $\mathrm{S}_{0}$-$\mathrm{S}_{3}$
and $\mathrm{S}_{0}$-$\mathrm{S}_{4}$.}
\end{figure*}

\textit{Excitation of dark state of ZnPc} -- To show its capability
in practical applications, we study the excitation of a zinc-phthalocyanine
(ZnPc) molecule, widely used in the STML experiments \citep{YangZhang_Nature_2016_dipole_dipole,Doppagne_science_2018}.
Tab. (\ref{tab:Table}) shows the transition dipoles and the energy
gaps of  the ZnPc molecule used in this paper. Due to the $D_{4h}$
symmetry of ZnPc (see the inset in Fig. \ref{fig:schematic_diagrame}
(a)), its first two excited states ($\mathrm{S}_{1}$ and $\mathrm{S}_{2}$)
whose eigenenergy is $1.98$eV are doubly degenerate. Both the $\mathrm{S}_{1}$
and $\mathrm{S}_{2}$ states, i.e., the Q states \citep{1970_Q_band,2001_Q_band},
are bright states, and their excitation and luminescence have already
been observed in experiments. Its third and fourth eigenstates $\mathrm{S}_{3}$
and $\mathrm{S}_{4}$ with eigenenergy $3.69$eV are degenerate dark
states. The details of these eigenstates and transition dipoles are
obtained with the time-dependent density functional theory (TDDFT)
at $\omega$B97X-D \citep{TDDFT_2008} /TZVP \citep{TDDFT_1994} level
by Gaussian 16 program \citep{Gaussian} and shown in Tab. (\ref{tab:Table}).
Fig. \ref{fig:ZnPc_tran=000026current}(a) shows the calculated transition
density of transition between the ground state $\mathrm{S}_{0}$ and
dark state $\mathrm{S}_{3}.$ The transition density is an even function
in the $y$-axis and an odd function in both the $x$- and $z$-axes.

Fig. \ref{fig:ZnPc_tran=000026current}(b) shows the 2D logarithmic
plot of the normalized inelastic current of $\mathrm{S}_{0}$-$\mathrm{S}_{3}$
transition (at $V_{\mathrm{b}}=-4$V). With the bias larger than the
energy of the $\mathrm{S}_{0}$-$\mathrm{S}_{3}$ transition, the
tunneling electron in STM allows the $\mathrm{S}_{0}$-$\mathrm{S}_{3}$
transition \citep{2020_GHDong_STML}. The 2D map of tunneling current
are obtained by moving the tip over the molecules at the constant
hight $d=1.0\mathrm{nm}$. The plot clearly shows the two  maxima
and four secondary  maxima in the transition density. With the same
bias, the transition $\mathrm{S}_{0}$-$\mathrm{S}_{4}$ is activated
simultaneously. The total current as the summation of the two transitions
$\mathrm{S}_{0}$-$\mathrm{S}_{3}$ and $\mathrm{S}_{0}$-$\mathrm{S}_{4}$
are shown  in Fig. \ref{fig:ZnPc_tran=000026current}(c). The profile
shows a four-lobe pattern which is similar to that of the bright state
observed in the experiment \citep{YangZhang_Nature_2016_dipole_dipole}.

Unlike the excitation of the molecular bright state, the molecule
in its dark state can not decay to its ground state through spontaneous
emission due to the optical selection rule. The lifetime of the dark
state is much longer than that of the bright state. The inelastic
current induced by the dark-state excitation approaches zero when
the dark state is totally excited (the population of dark state is
unity). The stable excitation of the dark state can be obtained with
a designed cyclic scheme as follows.

\textit{Detection of dark-state excitation of ZnPc} -- One possible
approach is to excite the dark state to a higher bright state and
detect the luminescence of the bright state. To illuminate our proposal,
we choose the higher bright state $\mathrm{S}_{17}$ with eigenenergy
$4.45$eV. The transition dipole of $\mathrm{S}_{0}$-$\mathrm{S}_{17}$
transition has a nonzero $z$-component (-0.21 a.u.), and that of
$\mathrm{S}_{3}$-$\mathrm{S}_{17}$ transition has a nonzero $x$-component
(0.27 a.u.). Both the two transitions are optically allowed. As shown
in Fig. \ref{fig:schematic_diagrame} (b), the molecule in its ground
state is excited to the dark state through the IES process with STM.
A laser at the wavelength $1635.7$nm ($0.76$eV) pumps the molecule
resonantly from the state $\mathrm{S}_{3}$ to state $\mathrm{S}_{17}$.
ZnPc in state $\mathrm{S}_{17}$ will emit photons at two wavelength
$1635.7$nm and $279.4$nm (the $\mathrm{S}_{0}$-$\mathrm{S}_{17}$
transition). The luminescence photons are collected at the wavelength
$279.4$nm. The kinetic equations of the populations on the three
states are written as
\begin{align}
\dot{P}_{0}\left(t\right) & =-\frac{I_{ine,03}}{e}P_{0}\left(t\right)+\gamma_{0}P_{17}\left(t\right),\nonumber \\
\dot{P}_{3}\left(t\right) & =-\eta P_{3}\left(t\right)+\frac{I_{ine,03}}{e}P_{0}\left(t\right)+\gamma_{3}P_{17}\left(t\right),\\
\dot{P}_{17}\left(t\right) & =-\left(\gamma_{0}+\gamma_{3}\right)P_{17}\left(t\right)+\eta P_{3}\left(t\right),\nonumber
\end{align}
where $I_{ine,03}$ is the inelastic current of the $\mathrm{S}_{0}$-$\mathrm{S}_{3}$
transition and $\eta$ characterizes the transition pump rate induced
by the pumping laser. And $\gamma_{0}(\gamma_{3})$ is the spontaneous
emission rate from the state $\mathrm{S}_{17}$ to the state $\mathrm{S}_{0}$($\mathrm{S}_{3}$).
In the steady state, the photon emission rate from $\mathrm{S}_{17}$
to $\mathrm{S}_{0}$ is
\begin{equation}
\Gamma=\gamma_{0}P_{17,\mathrm{s}}=\gamma_{0}\frac{I_{ine,03}}{e}\left(\gamma_{0}+\frac{I_{ine,03}}{e}\frac{\gamma_{0}+\gamma_{3}+\eta}{\eta}\right)^{-1},\label{eq:Gamma}
\end{equation}
where $P_{17,\mathrm{s}}$ is the population of state $\mathrm{S}_{17}$
in the steady state.

The photon emission rate in Eq. (\ref{eq:Gamma}) approximately equals
the inelastic current over an electron charge $I_{ine,03}/e$. In
the STML experiment, the photon yield (luminescence probability) is
as small as $10^{-5}$ photon/electron \citep{MichaelChong_thesis_2016}.
For the ZnPc molecule, the total excitation rate ($I_{ine}/e$) is
estimated approximately as $1.3\times10^{4}\mathrm{s}^{-1}$ \citep{YangZhang_Nature_2016_dipole_dipole,2020_GHDong_STML}.
As a result, the dark-state excitation rate $I_{ine,03}/e$ induced
by the IES process should be several orders of magnitude smaller than
$1.3\times10^{4}\mathrm{s}^{-1}$. For a moderate laser pump ($\eta\gg\gamma_{0}+\gamma_{3}$),
the second term in the parenthesis of Eq. (\ref{eq:Gamma}) is approximately
$I_{ine,03}/e$. The emission rate of the $\mathrm{S}_{17}$-$\mathrm{S}_{0}$
transition reads $\gamma_{0}=4\times10^{4}\mathrm{s}^{-1}$ which
will be much larger than the second term.

\textit{Conclusion} -- We propose a new perspective for STM to profile
the dark-state transition density  and demonstrate its capability
in both the proof-of-principle example and  the simulation of the
practical application with the ZnPc molecule. Benefiting from the
sub-nanometer resolution, STM can excite the molecular dark state
beyond the dipole approximation and the inelastic current inherits
the main characters of its corresponding transition density in the
sub-molecular scale. The additional laser pump to the bright state
allows the observation of the characteristic features in the current
with photon counting. The current proposal will extend the application
of STM to probe the photoprotection and energy-transfer effect on
the single molecule level.
\begin{acknowledgments}
H.D. thanks the support from the NSFC (Grant No. 11875049), the NSAF
(Grants No. U1730449 and No. U1930403), and the National Basic Research
Program of China (Grant No. 2016YFA0301201). X.S. thanks the support
from the NSFC (Grant No. 21903054).
\end{acknowledgments}

\bibliographystyle{apsrev4-1}
\bibliography{reference}

\end{document}


\begin{center}
Supplementary Material for
\par\end{center}

\title{Directly profiling the dark-state transition density via scanning
tunneling microscope}

\author{Guohui Dong}

\affiliation{Graduate School of Chinese Academy of Engineering Physics, Beijing
100084, China}

\author{Zhubin Hu}

\affiliation{New York University Shanghai and NYU-ECNU Center for Computational
Chemistry, 1555 Century Avenue, Shanghai 200122, China}

\affiliation{Department of Chemistry, New York University, New York, New York
10003, United States }

\author{Xiang Sun}
\email{xiang.sun@nyu.edu}

\affiliation{New York University Shanghai and NYU-ECNU Center for Computational
Chemistry, 1555 Century Avenue, Shanghai 200122, China}

\author{Hui Dong}
\email{hdong@gscaep.ac.cn}

\affiliation{Graduate School of Chinese Academy of Engineering Physics, Beijing
100084, China}

\maketitle
This document is devoted to providing the detailed derivations and
the supporting discussions in the main content.

\section{The electron-molecule interaction}

Under the single-electron approximation, the coupling between the
tunneling electron and the molecule is written as
\begin{equation}
\hat{H}_{\mathrm{el-m}}=\sum_{p=1}^{n}\frac{Z_{p}e^{2}}{\left|\hat{\vec{r}}_{0}-\hat{\vec{R}}_{p}\right|}-\sum_{q=1}^{2N}\frac{e^{2}}{\left|\hat{\vec{r}}_{0}-\hat{\vec{r}}_{q}\right|},
\end{equation}
where the first term describes the Coulomb interaction between the
nucleus (vector $\vec{R}_{p}$) and the tunneling electron (vector
$\vec{r}_{0}$), and the second term shows the interaction between
the molecular electron (vector $\vec{r}_{q}$) and the tunneling electron.
$n$ denotes the total number of atomic nuclei and $2N$ is that of
the electron ($2N=\sum_{p=1}^{n}Z_{p}$). 
\begin{figure}
\begin{centering}
\includegraphics{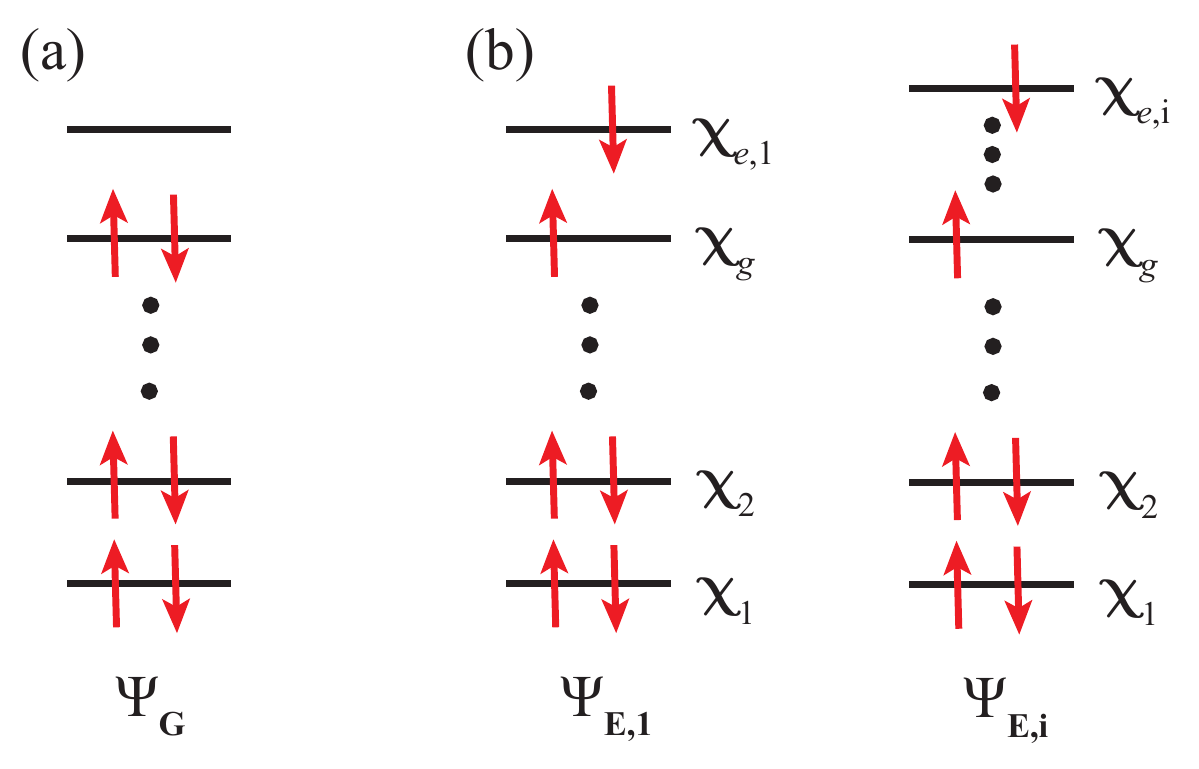}
\par\end{centering}
\caption{\label{fig:Orbital-energy-level-diagram}Orbital energy-level diagram
for electronic configuration of (a) ground state $\Psi_{G}$ and (b)
excited state $\Psi_{E,i}$. }
\end{figure}
 With the antisymmetric property of electrons, we assume that the
 many-electron wavefunction of the molecular ground and excited states
are written as (Fig. \ref{fig:Orbital-energy-level-diagram})
\begin{align*}
\Psi_{G}\left(1,2,...2N\right) & =\Pi\sum_{k}\left(-1\right)^{P}\hat{P}_{k}\left[\chi_{1,\uparrow}\left(1\right)\chi_{1,\downarrow}\left(2\right)\chi_{2,\uparrow}\left(3\right)\chi_{2,\downarrow}\left(4\right)...\chi_{g,\uparrow}\left(2N-1\right)\chi_{g,\downarrow}\left(2N\right)\right],\\
\Psi_{E,i}\left(1,2,...2N\right) & =\Pi\sum_{k}\left(-1\right)^{P}\hat{P}_{k}\left[\chi_{1,\uparrow}\left(1\right)\chi_{1,\downarrow}\left(2\right)\chi_{2,\uparrow}\left(3\right)\chi_{2,\downarrow}\left(4\right)...\chi_{g,\uparrow}\left(2N-1\right)\chi_{e,i,\downarrow}\left(2N\right)\right],
\end{align*}
where $\hat{P}_{k}$ is the $k$-th permutation operators, $\left(-1\right)^{P}$
is +1 for an even permutation and -1 for an odd permutation, the summation
contains all the permutations, and  $\Pi=1/\sqrt{\left(2N\right)!}$
is the normalization factor.

In the sub-space of the ground state $\Psi_{G}$ and excited state
$\left|\Psi_{E,i}\right\rangle \left(i=1,\cdots,l\right)$, we find
the coupling term as 
\begin{align}
\hat{H}_{\mathrm{el-m}} & =\left(\left|\Psi_{G}\right\rangle \left\langle \Psi_{G}\right|+\sum_{i=1}^{l}\left|\Psi_{E,i}\right\rangle \left\langle \Psi_{E,i}\right|\right)\left(\sum_{p=1}^{n}\frac{Z_{n}e^{2}}{\left|\hat{\vec{r}}_{0}-\hat{\vec{R}}_{p}\right|}-\sum_{q=1}^{2N}\frac{e^{2}}{\left|\hat{\vec{r}}_{0}-\hat{\vec{r}}_{q}\right|}\right)\left(\left|\Psi_{G}\right\rangle \left\langle \Psi_{G}\right|+\sum_{j=1}^{l}\left|\Psi_{E,j}\right\rangle \left\langle \Psi_{E,j}\right|\right)\nonumber \\
 & =\sum_{p=1}^{n}\frac{Z_{n}e^{2}}{\left|\hat{\vec{r}}_{0}-\hat{\vec{R}}_{p}\right|}-\sum_{q=1}^{2N}\left(\sum_{i,j=1}^{l}\left\langle \Psi_{E,i}\right|\frac{e^{2}}{\left|\hat{\vec{r}}_{0}-\hat{\vec{r}}_{q}\right|}\left|\Psi_{E,j}\right\rangle \left|\Psi_{E,i}\right\rangle \left\langle \Psi_{E,j}\right|+\left\langle \Psi_{G}\right|\frac{e^{2}}{\left|\hat{\vec{r}}_{0}-\hat{\vec{r}}_{q}\right|}\left|\Psi_{G}\right\rangle \left|\Psi_{G}\right\rangle \left\langle \Psi_{G}\right|\right)\nonumber \\
 & -\frac{2N}{\left(2N\right)!}\sum_{i=1}^{l}\left(\left\langle \Psi_{E,i}\right|\frac{e^{2}}{\left|\hat{\vec{r}}_{0}-\hat{\vec{r}}_{1}\right|}\left|\Psi_{G}\right\rangle \left|\Psi_{E,i}\right\rangle \left\langle \Psi_{G}\right|+\left\langle \Psi_{G}\right|\frac{e^{2}}{\left|\hat{\vec{r}}_{0}-\hat{\vec{r}}_{1}\right|}\left|\Psi_{E,i}\right\rangle \left|\Psi_{G}\right\rangle \left\langle \Psi_{E,i}\right|\right)\nonumber \\
 & =\sum_{p=1}^{n}\frac{Z_{n}e^{2}}{\left|\hat{\vec{r}}_{0}-\hat{\vec{R}}_{p}\right|}-\sum_{q=1}^{2N}\left(\sum_{i,j=1}^{l}\left\langle \Psi_{E,i}\right|\frac{e^{2}}{\left|\hat{\vec{r}}_{0}-\hat{\vec{r}}_{q}\right|}\left|\Psi_{E,j}\right\rangle \left|\Psi_{E,i}\right\rangle \left\langle \Psi_{E,j}\right|+\left\langle \Psi_{G}\right|\frac{e^{2}}{\left|\hat{\vec{r}}_{0}-\hat{\vec{r}}_{q}\right|}\left|\Psi_{G}\right\rangle \left|\Psi_{G}\right\rangle \left\langle \Psi_{G}\right|\right)\nonumber \\
 & -\frac{2N}{\left(2N\right)!}\sum_{i=1}^{l}\left(\left(2N-1\right)!\left\langle \chi_{e,i}\right|\frac{e^{2}}{\left|\hat{\vec{r}}_{0}-\hat{\vec{r}}_{1}\right|}\left|\chi_{g}\right\rangle \left|\Psi_{E,i}\right\rangle \left\langle \Psi_{G}\right|+\left(2N-1\right)!\left\langle \chi_{g}\right|\frac{e^{2}}{\left|\hat{\vec{r}}_{0}-\hat{\vec{r}}_{1}\right|}\left|\chi_{e,i}\right\rangle \left|\Psi_{G}\right\rangle \left\langle \Psi_{E,i}\right|\right)\nonumber \\
 & =\sum_{p=1}^{n}\frac{Z_{n}e^{2}}{\left|\hat{\vec{r}}_{0}-\hat{\vec{R}}_{p}\right|}-\sum_{q=1}^{2N}\left(\sum_{i,j=1}^{l}\left\langle \Psi_{E,i}\right|\frac{e^{2}}{\left|\hat{\vec{r}}_{0}-\hat{\vec{r}}_{q}\right|}\left|\Psi_{E,j}\right\rangle \left|\Psi_{E,i}\right\rangle \left\langle \Psi_{E,j}\right|+\left\langle \Psi_{G}\right|\frac{e^{2}}{\left|\hat{\vec{r}}_{0}-\hat{\vec{r}}_{j}\right|}\left|\Psi_{G}\right\rangle \left|\Psi_{G}\right\rangle \left\langle \Psi_{G}\right|\right)\nonumber \\
 & -\sum_{i=1}^{l}\int d^{3}\vec{r}\frac{e^{2}\rho_{T,i}\left(\vec{r}\right)}{\left|\hat{\vec{r}}_{0}-\vec{r}\right|}\left(\left|\Psi_{E,i}\right\rangle \left\langle \Psi_{G}\right|+\left|\Psi_{G}\right\rangle \left\langle \Psi_{E,i}\right|\right),\label{eq:He_m}
\end{align}
where we have defined transition density $\rho_{T,i}\left(\vec{r}\right)=\chi_{e,i}^{*}\left(\vec{r}\right)\chi_{g}\left(\vec{r}\right)=\chi_{g}^{*}\left(\vec{r}\right)\chi_{e,i}\left(\vec{r}\right)$
\citep{transition_density_2008}. 

The last term in Eq. (\ref{eq:He_m}) represents the molecular transition
between its ground and excited states. In the single-electron Hilbert
space, we rewrite the last term of the interaction Hamiltonian as
\[
\hat{H}_{\mathrm{el-m}}\simeq-\sum_{i=1}^{l}\int d^{3}\vec{r}\frac{e^{2}\rho_{T,i}\left(\vec{r}\right)}{\left|\hat{\vec{r}}_{0}-\vec{r}\right|}\left(\left|\chi_{e,i}\right\rangle \left\langle \chi_{g}\right|+\left|\chi_{g}\right\rangle \left\langle \chi_{e,i}\right|\right)
\]
Thus, in the main text we focus on the calculation of this term. The
transition matrix element of transition $\left|\chi_{e,i}\right\rangle \left|\varphi_{n}\right\rangle \rightarrow\left|\chi_{g}\right\rangle \left|\phi_{k_{0}}\right\rangle $
(negative bias) is 
\begin{align}
\mathscr{N}_{\mathrm{s,t};i}|_{-V_{\mathrm{b}},E_{n}\rightarrow\xi_{k_{0}}} & =\int d^{3}\vec{r}_{0}\int d^{3}\vec{r}\frac{\rho_{T,i}\left(\vec{r}\right)e^{2}}{\left|\vec{r}_{0}-\vec{r}\right|}\varphi_{n}\left(\vec{r}_{0}\right)\phi_{k_{0}}\left(\vec{r}_{0}\right)\nonumber \\
 & =\int d^{3}\vec{r}_{0}\int d^{3}\vec{r}\frac{\rho_{T,i}\left(\vec{r}\right)e^{2}}{\left|\vec{r}_{0}-\vec{r}\right|}\rho_{n,k_{0}}\left(\vec{r}_{0}\right)\nonumber \\
 & =\frac{1}{\left(2\pi\right)^{3}}\int d^{3}\vec{r}_{0}\int d^{3}\vec{r}\int\int d^{3}\vec{p}d^{3}\vec{q}\frac{\rho_{T,i}\left(\vec{q}\right)e^{2}}{\left|\vec{r}_{0}-\vec{r}\right|}\rho_{n,k_{0}}\left(\vec{p}\right)e^{-i\vec{p}\cdot\left(\vec{r}_{0}-\vec{r}\right)}e^{-i\left(\vec{p}+\vec{q}\right)\cdot\vec{r}}\nonumber \\
 & =\frac{1}{\left(2\pi\right)^{3}}\int d^{3}\vec{r}_{0}\int d^{3}\vec{r}\int\int d^{3}\vec{p}d^{3}\vec{q}\frac{\rho_{T,i}\left(\vec{q}\right)e^{2}}{\left|\vec{r}_{0}\right|}\rho_{n,k_{0}}\left(\vec{p}\right)e^{-i\vec{p}\cdot\vec{r}_{0}}e^{-i\left(\vec{p}+\vec{q}\right)\cdot\vec{r}}\nonumber \\
 & =\int d^{3}\vec{r}_{0}\int\int d^{3}\vec{p}d^{3}\vec{q}\frac{\rho_{T,i}\left(\vec{q}\right)e^{2}}{\left|\vec{r}_{0}\right|}\rho_{n,k_{0}}\left(\vec{p}\right)e^{-i\vec{p}\cdot\vec{r}_{0}}\delta\left(\vec{p}+\vec{q}\right)\nonumber \\
 & =\int d^{3}\vec{q}\rho_{T,i}\left(\vec{q}\right)\rho_{n,k_{0}}\left(-\vec{q}\right)\left[\int d^{3}\vec{r}\frac{e^{2}}{\left|\vec{r}\right|}e^{i\vec{q}\cdot\vec{r}}\right]\nonumber \\
 & =e^{2}\int d^{3}\vec{q}\rho_{T,i}(\vec{q})\rho_{n,k_{0}}(-\vec{q})\frac{4\pi}{q^{2}},
\end{align}
where we have used the notation $\rho_{n,k_{0}}\left(\vec{r}\right)=\varphi_{n}\left(\vec{r}\right)\phi_{k_{0}}\left(\vec{r}\right)$,
and $\rho_{T,i}(\vec{k}),\rho_{n,k_{0}}(\vec{k})$ are the Fourier
transformations of $\rho_{T,i}(\vec{r}),\rho_{n,k_{0}}(\vec{r})$
\begin{align}
\rho_{T,i}(\vec{k}) & =\frac{1}{\sqrt{2\pi}^{3}}\int d^{3}\vec{r}\rho_{T,i}(\vec{r})e^{i\vec{k}\cdot\vec{r}},\\
\rho_{n,k_{0}}(\vec{k}) & =\frac{1}{\sqrt{2\pi}^{3}}\int d^{3}\vec{r}\varphi_{n}\left(\vec{r}\right)\phi_{k_{0}}\left(\overrightarrow{r}\right)e^{i\vec{k}\cdot\vec{r}}.
\end{align}

\section{current at positive bias}

For the positive bias, the inelastic current has a similar form which
is 
\begin{align}
I_{+\mathrm{,inela}} & \simeq\sum_{i=1}^{l}I_{+\mathrm{,inela},i},\label{eq:inela_current-1}
\end{align}
where 

\begin{align}
I_{+\mathrm{,inela},i} & =2\pi e\int_{\mu_{0}}^{\mu_{0}+eV_{b}-E_{eg,i}}dE_{n}\rho_{\mathrm{s}}\left(E_{n}\right)\rho_{\mathrm{t}}\left(\xi_{k}\right)\nonumber \\
 & \times\left|\mathscr{N}_{\mathrm{t,s};i}|_{V_{\mathrm{b}},\xi_{k}\rightarrow E_{n}}\right|^{2}|_{\xi_{k}=E_{n}-eV_{\mathrm{b}}+E_{eg,i}}.\label{eq:inela_current_i-1}
\end{align}

\section{Asymmetry of current}

\begin{figure}
\begin{centering}
\includegraphics[scale=0.5]{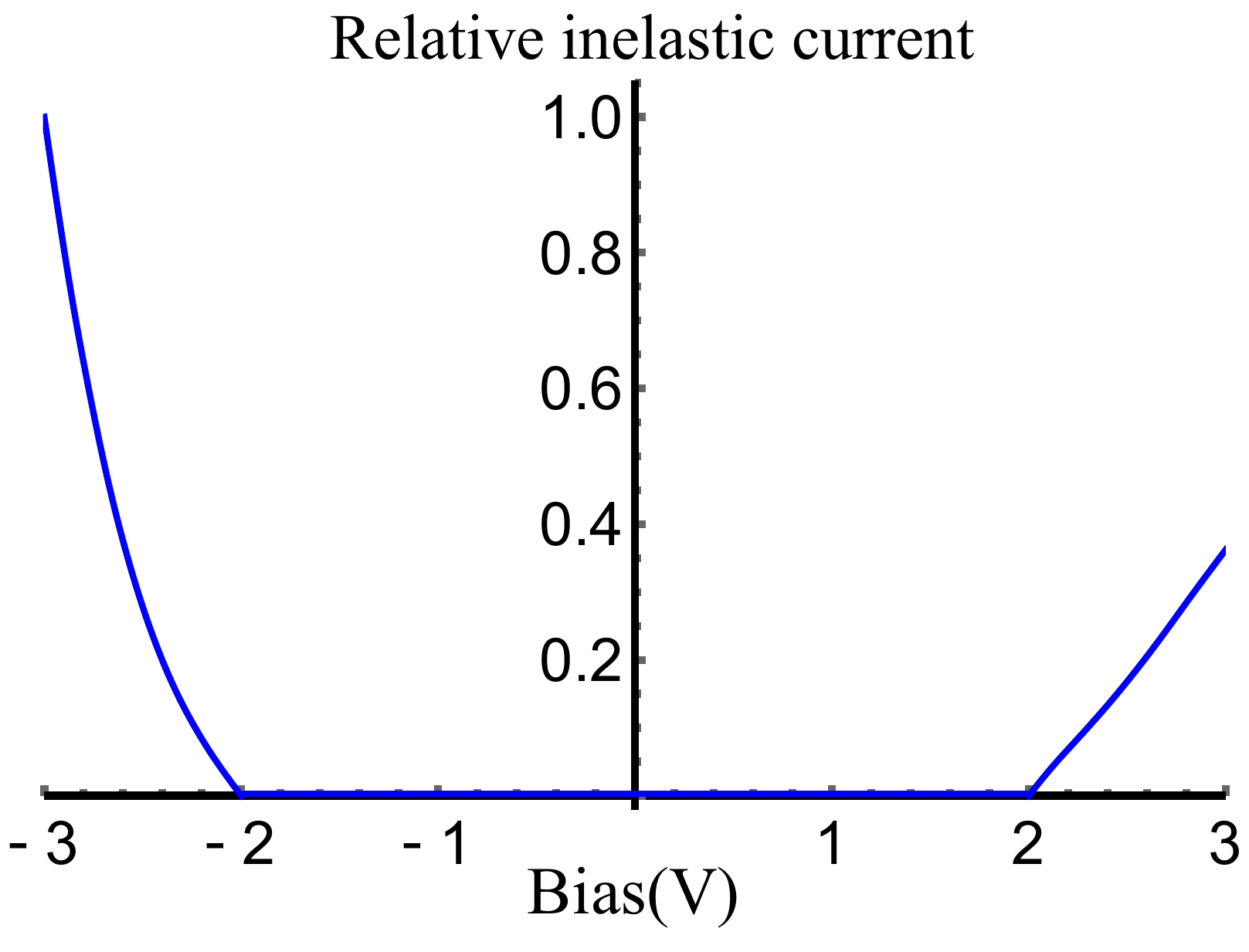}
\par\end{centering}
\caption{\label{fig:Gaussian_result_asymmetry} (Color online) The relative
inelastic current when the tip is place right above point $\left(0,0\right)$.
Here we choose $2\sigma_{1}=\sigma_{2}=\sigma=1$nm, $R=0.5$nm, $d=1$nm,
and $E_{eg}=2$eV. }
\end{figure}

We study the inelastic current of the Gaussian transition density
for different bias voltage. Fig. \ref{fig:Gaussian_result_asymmetry}
depicts the varies of the inelastic current as a function of bias
voltage. Here we choose that both the tip and substrate are made of
silver, and the Fermi energy is $\mu_{0}=-4.64$eV. The radius of
the tip is $R=0.5$nm, and the distance between the molecular plane
and tip is $d=1$nm. The energy gap between the molecular ground state
and dark excited state is $E_{eg}=2$eV. Without the loss of generality,
we  choose the tip to be right above the center of the molecular transition
density, namely $A=(0,0)$. Fig. \ref{fig:Gaussian_result_asymmetry}
shows that when the bias voltage is larger than the molecular energy
gap ($\left|eV_{\mathrm{b}}\right|>E_{eg}$), the inelastic current
becomes nonzero. Moreover, the inelastic current at negative bias
is larger than that at positive bias. Actually, this two features
are also found in the study of inelastic electron scattering mechanism
under the dipole approximation \citep{2020_GHDong_STML}.  The two
features are typical in the IES mechanism in the cases with or without
the dipole approximation.

\section{Excitation of ZnPc bright state}

As shown in the main text, the total inelastic current of the system
consists of the contribution of all the transitions whose transition
energy is lower than $eV_{\mathrm{b}}$. Without loss of generality,
here we choose a negative bias $V_{\mathrm{b}}=-2.5$V. At $V_{\mathrm{b}}=-2.5$V,
the total inelastic current is the sum of the current induced by transitions
$\mathrm{S}_{0}$-$\mathrm{S}_{1}$ and $\mathrm{S}_{0}$-$\mathrm{S}_{2}$.

\begin{figure}[h]
\begin{centering}
\includegraphics[scale=0.25]{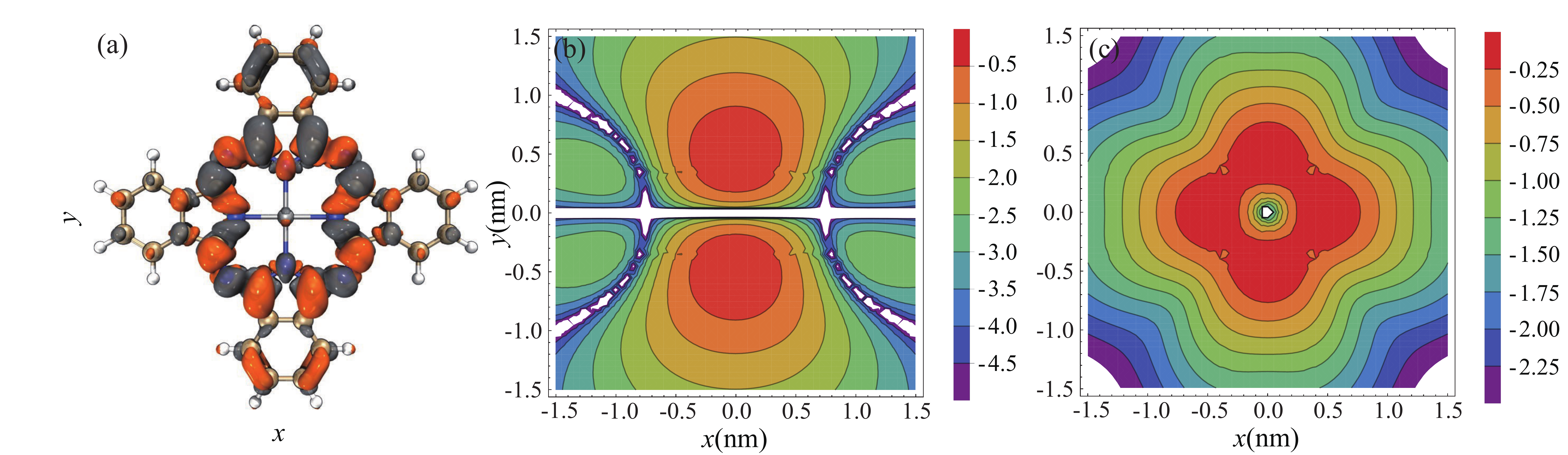}
\par\end{centering}
\centering{}\caption{\label{fig:ZnPc_Bright} (a) The transition density of transition
between the ground state $\mathrm{S}_{0}$ and bright state $\mathrm{S}_{1}$.
(b) The 2D logarithmic plot of the relative inelastic current of transition
between the $\mathrm{S}_{0}$ and $\mathrm{S}_{1}$ at $V_{\mathrm{b}}=-2.5$V.
(c) The 2D logarithmic plot of the sum of the relative inelastic current
of transitions $\mathrm{S}_{0}$-$\mathrm{S}_{1}$ and $\mathrm{S}_{0}$-$\mathrm{S}_{2}$. }
\end{figure}

Fig. \ref{fig:ZnPc_Bright}(a) depicts the transition density of transition
between the ground state $\mathrm{S}_{0}$ and bright state $\mathrm{S}_{1}$.
The transition density is an odd function in the $y$-axis and an
even function in both the $x$- and $z$-axes, which gives a nonzero
transition dipole moment in the $y$-axis. Fig. \ref{fig:ZnPc_Bright}(b)
plots the 2D logarithmic plot of the normalized inelastic current
(luminescence strength) of $\mathrm{S}_{0}$-$\mathrm{S}_{1}$ transition.
As the transition dipole lies along the $y$-axis, the inelastic current
gives a two-spot pattern along the $y$-axis. The inelastic current
(luminescence strength) around the $x$-axis is smaller than that
along the $y$-axis by several orders of magnitude and thus is undetectable.
Due to the symmetry of the ZnPc molecule, the transition density of
$\mathrm{S}_{0}$-$\mathrm{S}_{2}$ transition can be obtained by
rotating the Fig. \ref{fig:ZnPc_Bright}(a) $90^{\mathrm{o}}$ along
the $z$-axis. The total inelastic current (luminescence strength)
contributed by $\mathrm{S}_{0}$-$\mathrm{S}_{1}$ and $\mathrm{S}_{0}$-$\mathrm{S}_{2}$
transition is shown in Fig. \ref{fig:ZnPc_Bright}(c). We can see
that the total current inherits the four-lobe pattern from the geometric
symmetry of the ZnPc, which has been observed in experiments \citep{YangZhang_Nature_2016_dipole_dipole,YangLuo_Single_photon_superradiance_PRL_2019}.

\bibliographystyle{apsrev4-1}
\bibliography{reference}